\documentclass[prl, twocolumn, footinbib,a4paper, showpacs]{revtex4}
\usepackage[latin1]{inputenc}
\usepackage{amssymb}
\usepackage{amsmath}
\usepackage{amsbsy}
\usepackage{bm}
\usepackage{graphicx}

\begin{document}
\title{ Room temperature water Leidenfrost droplets  \\ }

\author{Franck Celestini $^1$, Thomas Frisch$^2$ , Yves Pomeau$^3$ }

\affiliation{$^1$ Laboratoire de Physique de la Mati\`ere Condens\'ee, CNRS UMR 7366, Universit\'e de Nice Sophia-Antipolis,
   Parc Valrose  06108 Nice Cedex 2, France}
   
\affiliation{$^2$ Institut  Non Lin\'eaire de Nice, CNRS UMR 7735,  Universit\'e de Nice Sophia-Antipolis,
   1361 Routes des lucioles, Sophia Antipolis F-06560 Valbonne France}

\affiliation{$^3$  University of Arizona, Department of Mathematics, Tucson, AZ 85721 USA}

\begin{abstract}
We experimentally investigate the Leidenfrost effect at pressures ranging from $1$ to $0.05$ atmospheric pressure. As a direct consequence of the Clausius-Clapeyron phase diagram of water,  the droplet temperature can be at ambient temperature in a non-sophisticated lab environment. Furthermore, the lifetime of the Leidenfrost droplet is significantly increased in this low pressure environment. The temperature and pressure dependance of the evaporation rate are successfully tested against a recently proposed model. These results may pave  a way to reach efficient Leidenfrost micro-fluidic and milli-fluidic  applications.
\end{abstract}


\maketitle

  There has been lately numerous studies of the {\it{dynamical}} properties of  Leidenfrost droplets \cite{ PRL,chicago, Celestini, linke, quere, gold, xu2013}. Undoubtedly one of the reasons of this interest is in  the possible applications in micro-fluidic and milli-fluidic.  These droplets   float on a   thin vapor  film which strongly delays their evaporation rate due  to the low thermal heat conductivity of the vapor as compared to the one of the liquid. The transport by Leidenfrost droplets has the big advantage, compared to regular microfluidic in small pipes, to be without contact with  solid surfaces  which are oftenly  a potential source of pollution.  However this advantage of no contact with the surfaces  is plagued by other inconveniences. Among them, the temperature inside  water Leidenfrost droplets  is usually too high and is  in the vicinity  of the boiling temperature of the liquid they are made of.  In particular,  the temperature water   near boiling at atmospheric room pressure is  often  far above the range of stability of most biomolecules which could be contained in the droplet.
   The investigation we report below is devoted to a simple idea, namely the one of using  a  {\it  low pressure environment} to lower the boiling temperature to the room temperature.
This may pave the way to potential micro-fluidics and milli-fluidic  devices using Leidenfrost  droplets as small  room temperature reactors. Since the boiling temperature of water falls rather sharply as pressure decreases, it is in principle not difficult to reach pressures where most biomolecules would survive in Leidenfrost droplets. However, since the Leidenfrost effect on droplets is a rather complex phenomenon \cite{CRAS}, we found it useful to check that it still exists at low pressure and low  temperatures with water droplets.  
 Moreover, the controlled formation and smooth release of small spherical droplets in low pressure  situations is not such a trivial task, at least without recourse to complex laboratory methods.  \\
 We present  in this Letter  a realization  using fairly standard methods which are likely easily reproducible in a non too sophisticated lab environment.  We first  put forth our experimental set-up  and the results obtained for the evolution of the droplet  radius  as function of  time in low pressure conditions.
 We experimentally examine the mass evaporation rate as a function of the applied pressure $P$  and of the temperature difference between the Leidenfrost droplet   and the heated  plate on which it is levitating. We put in evidence   a significant increased lifetime of a  low pressure  Leidenfrost droplet. These results are successfully compared with the recently proposed  theoretical analysis  of the Leidenfrost phenomena  \cite{CRAS}. We finally conclude this paper by pointing out the possible applications of this  room temperature Leidenfrost effect.

\begin{figure}[!h]
 \includegraphics[width=.5\textwidth]{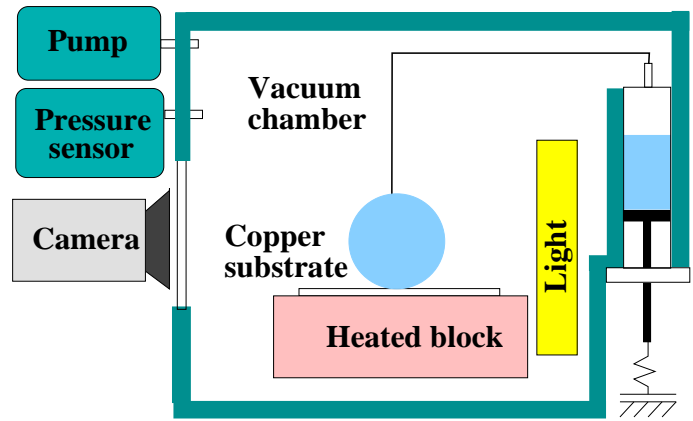}
 \caption{Sketch of the experimental set-up. }
 \label{fig1}
\end{figure}

 The experimental set-up is displayed in Fig. \ref{fig1}.  A primary vacuum pump is used to  impose   the pressure $P$ in the chamber of  volume  $10^{-2} \,  \rm{m}^3$.
 A differential pressure sensor measures accurately the pressure $P$ in the $1000$ - $1$ mbar range. The droplets are deposited on the substrate with a syringe inserted in the vacuum  chamber  in a vertical position. This last point is of importance since it permits to evacuate  gas bubbles dissolved  in the water which appear when the pressure is lowered. The temperature of the droplet is denoted $T$.   As for the classical (atmospheric pressure)  Leidenfrost effect,
the liquid is at thermal equilibrium with the vapor. As a consequence the value of  $T$ follows from the pressure measurement and the classical water liquid-vapor phase diagram (P-T).  
  The substrate is a polished copper block heated at a controlled temperature $T_s$ and the temperature difference between the heated plate and the droplet  is denoted $\Delta T = T_s-T$. The substrate temperature is directly measured using a temperature sensor  inserted in the heated block.
   A glass window  in the vacuum chamber permits  the observation of the droplet with a camera. Once the droplet is deposited on the substrate,  we record the decrease of its  radius $R(t)$ due to the evaporation on the heated substrate.  We  deduce  the volume of the droplet  and the mass evaporation rate  $J \propto \rho_l  \frac{\partial R^3}{\partial t}$.   Here we will focus on the results obtained for droplet with radius lower than the capillary length $R_c=(\frac{\sigma}{\rho_l g})^{1/2}$,  $\sigma$ being the surface tension between the liquid and the vapor, $\rho_l$ the mass density of the liquid  and $g$ the acceleration of gravity. 
  
  \begin{figure}[!h]
\includegraphics[width=.40\textwidth]{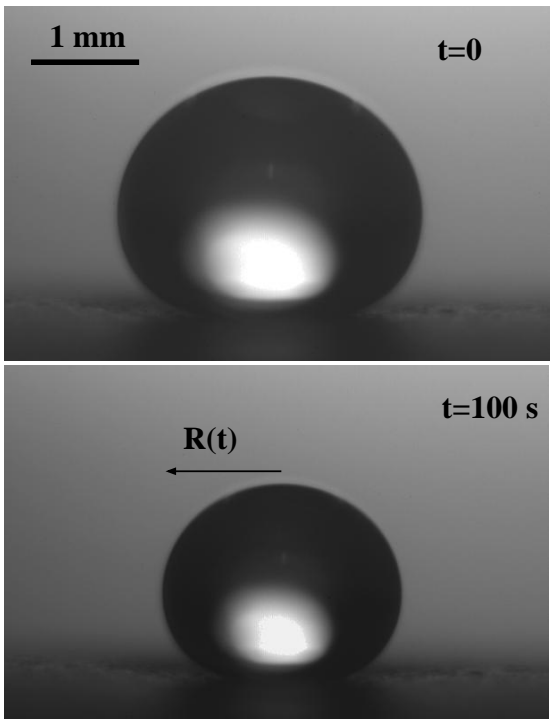}
 \caption{  View from the side of the Leidenfrost droplet for $t=0  \, s $ and $t=100 \,  s $ at a pressure of 1/20 atmosphere ($50  \,  {\rm mbar}$) . At this pressure, the temperature of the Leidenfrost  droplet  is $33  \,  {\rm  Celsius}$. }
 \label{fig2}
 \end{figure}
  
\begin{figure}
\includegraphics[width=.40\textwidth]{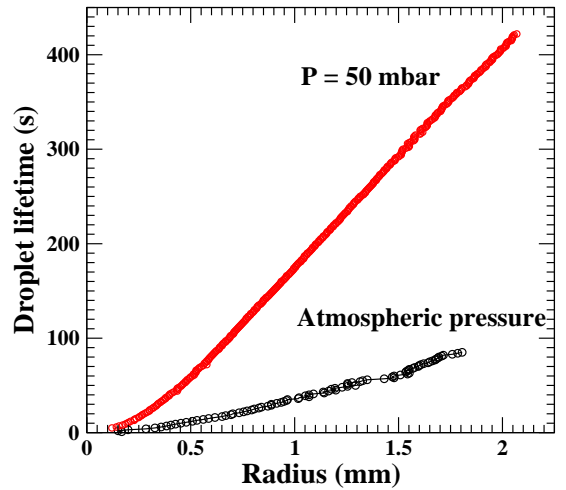}
\caption{ Leidenfrost droplet life-time as function of the radius for   two different  pressure:  top red (color on-line)  $P=50  \, {\rm mbar}  $, bottom black  $P=1 \,   {\rm bar} $.   The temperature of the  substrate is $120 \,  {\rm Celsius} $  for the top red  curve and  $250 \, {\rm Celsius} $  for the bottom black curve. }
\label{fig3}
\end{figure}

 As shown on  Fig. \ref{fig2},  we observe a  Leidenfrost droplet under a pressure of 50 mbar (1/20 atmosphere). It  has the shape of a weakly deformed sphere as expected for a droplet of radius smaller than the capillary length. This droplet floats on a  thin vapor film which has a small thermal conductivity compared to the one of the liquid.   The droplet is at a temperature $T$  corresponding to  the vapor-liquid equilibrium temperature at a given pressure. For the droplet represented in Fig. \ref{fig2}, the pressure  $P$ is  set to $50  \,  {\rm mbar}$  thus  leading to  a droplet  temperature of  $33  \,  {\rm  Celsius} $. As stressed above  the main characteristic of a Leidenfrost droplet under a  controlled pressure environment is that for sufficiently low pressure the drop can be at room temperature. Furthermore, another  attribute  of  the droplet is the    noticeable increase of  its   lifetime.   For a given initial radius $R$, this lifetime is simply defined as the time for the droplet to reach a radius that is no longer observable with our optical device. This  increase  is illustrated in Fig.  \ref{fig3}    for two different values of the pressure.  
  Indeed, the mass evaporation rate is related to the temperature difference between the droplet and the hot plate as expressed in equation (1) below.  The  temperature difference $\Delta T$ between the substrate  and the droplet is of  $87 \, {\rm Celsius} $   for the $50  \,  {\rm mbar}$  case and is of of $150  \, {\rm Celsius}$ for the   $1 \,  {\rm bar}$  case.  This smaller temperature gap at low pressure is a factor  which contributes to the  increases the life time of the droplet. As shown on Fig.  \ref{fig3} , one can   see that the lifetime of a {\it  low pressure Leidenfrost droplet}   is  increased by a factor of order $10$. For example, a droplet of initial radius  equal to the capillary length $R_c$  can  reach a lifetime of ten minutes.

This lifetime increase is not solely due to the pressure  reduction but also to the temperature difference $\Delta T$ between the plate and the drop. Let us recall the recently published  results for the mass evaporation rate  $J = d M/dt$ of Leidenfrost droplets \cite{CRAS}. For droplets of radii ranging between an ``intermediary length" $R_i=(R_l^3 R_c^4)^{1/7}$  \cite{note1}
  and the capillary length $R_c$  it has been  shown that $J$ scales as  :
\begin{equation}
J \propto R^{12/5} P^{1/5} \Delta T^{4/5} \,  .
\label{equation1}
\end{equation}
 This equation relates  the mass evaporation rate
to the radius of the droplet, to  the applied pressure and  to  the temperature difference between the droplet and the hot plate. Equation (1) has been derived from a theoretical analysis using the lubrication limit to model the vapor flow below the Leidenfrost droplet \cite{CRAS}.  It is worth noticing 
 that this prediction is   more elaborated  than the  standard approach of the
 Leidenfrost phenomenon  \cite{biance} that predicts $J \propto R \Delta T$ without any dependance on the 
 applied pressure. In Eq.(\ref{equation1}), the $P^{1/5}$ dependance  arises  from the  linear dependance  of  the vapor density  with 
  the pressure at constant temperature  \cite{note2}.

\begin{figure}
\includegraphics[width=.50\textwidth]{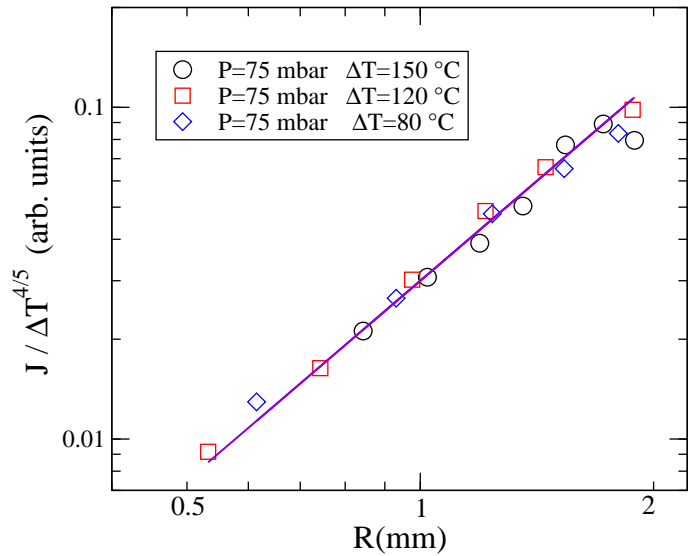}
 \caption{ Temperature normalized mass evaporation rate versus radius at constant pressure $p= 75  \, {\rm mbar}$ for three different  value of   $ \Delta T= 80,120,150  \,  {\rm  Celsius} $.
 The best fit  (full blue line, color on-line) leads to  $R^{2}$ tendency which  is close to the theoretical prediction of $R^{12/5}$.  \label{fig4}}
\end{figure}

To verify this prediction we first perform experiments to ascertain our scaling with respect to  $\Delta T$.  The pressure in the vacuum chamber is set to $75 \,  {\rm  mbar}$ and three evaporation rates are recorded for three different value of  the temperature difference:  $ \Delta T= 80,120 , 150  \, {\rm Celsius}   $.  As shown on  Fig. \ref{fig4},  we plot  $J/\Delta T^{4/5}$  as a function of the droplet radius. One can see that the data sets   collapse  on a same curve. Thus this strongly validates the  $\Delta T^{4/5}$ dependance of  $J$  given by Eq. (\ref{equation1}).  The best fit  of our experimental   data leads to  $J/  \Delta T^{4/5} \propto R^{2}$    which also lies in close   agreement with our prediction of $J/  \Delta T^{4/5} \propto R^{2.4}$.The relative agreement between the theory and the experiments is not surprising since the theory is made upon the lubrication approximation. A more accurate theoretical analysis would required the full resolution of Navier-Stokes for a biphasic system (liquid-vapor).
 It is important to note that this results hold for $R$ lying between $R_i$ and $R_c$ and that another scaling law may appear in different regimes ($R >R_c$ or $R_l < R < R_i$).

\begin{figure}
\includegraphics[width=.50\textwidth]{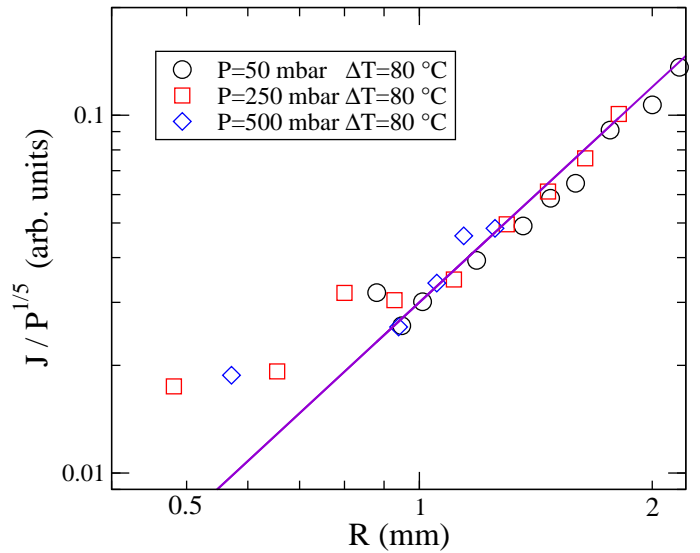}
 \caption{ Pressure  normalized mass evaporation rate versus radius at constant temperature $\Delta T = 80 \,  {\rm Celsius} $ for three different  values of the pressure : $P=50 \,  {\rm  mbar}$, $250  \, {\rm mbar} $ and $500  \,  {\rm  mbar}$. The best fit  (full blue line, color on-line) leads to  $R^{2}$ tendency which  is close the theoretical prediction of $R^{12/5}$  as in the previous figure.  \label{fi54}}
 \label{fig5}
\end{figure}
We now look at the dependance of the mass evaporation rate $J$ with  respect to the pressure. We  set  the value of $\Delta T$ to $80$ K and perform three experiments for different values of the pressure : $P=50 \,  {\rm  mbar}$, $250  \, {\rm mbar} $ and $500  \,  {\rm  mbar}$.  As shown  on Fig. \ref{fig5}, we plot  the pressure normalized mass evaporation rate  $J/P^{1/5}$ as a function of $R$. As for the previous analysis, these  data  collapse on the same curve, scaling as $R^{2}$,  evidencing a good agreement between the experiments and  theory.

We have demonstrated  in this  Letter that Leidenfrost droplets can be  generated  at  room temperature using a  non too sophisticated lab environment. The life-time of the droplets increases by roughly  one order of magnitude  compared to the standard atmospheric pressure conditions.  The  evaporation rates is decreased  at low pressure due to the small difference of temperature between the substrate and the droplet.  We have also verified that the scaling law for the mass evaporation rate as a function of the temperature, pressure and radii agree with  our recently proposed theory. We hope that this study will stimulate discussion and interest as a possible device in the micro and  milli-fluidic area in which hovering water droplet  at room temperature may serve as bio- or chemio-reactors \cite{kirby10,tabeling,bremond}.

\begin{acknowledgements}
 We would like to thank Jean-Pierre Romagnan and Patrick Tabeling for fruitful discussions.
 \end{acknowledgements}

  \end{document}